\documentclass{ws-ijmpd}
\usepackage[super,compress]{cite}
\usepackage{dcolumn}
\usepackage{bm}
\usepackage[hidelinks]{hyperref}
\usepackage{balance}
\DeclareMathOperator{\sech}{sech}

\renewcommand\Im{\operatorname{{Im}}}

\begin{document}

%title page
\title{An approach to the quantization of black hole quasi-normal modes}
\date{\today}

\author{Soham Pal\footnote{Currently at Iowa State University, Ames}}
%\email[]{E-mail: mssp2@iacs.res.in}
\address{ %Department of Materials Science, Indian Association for the Cultivation of Science, Kolkata 700032, India}
Department of Physics,\\ Birla Institute of Technology and Science, Pilani - Hyderabad Campus\\
Hyderabad 500078, India}
\author{Karthik Rajeev}
%\email[]{E-mail: karthikr@iisertvm.ac.in}
\author{ S.~Shankaranarayanan\footnote{Communicating author. Email: shanki@iisertvm.ac.in}}
\address{School of Physics,\\ Indian Institute of Science Education and 
Research (IISER-TVM),\\ Thiruvananthapuram 695016, India}

\maketitle

\begin{abstract}
	In this work we derive the asymptotic quasi-normal modes of a BTZ black hole using a quantum field theoretic Lagrangian. The BTZ black hole is a very popular system in the context of $2+1$-dimensional quantum gravity. However to our knowledge the quasi-normal modes of the BTZ black hole have been studied only in the classical domain. Here we show a way to quantize the quasi-normal modes of the BTZ black hole by mapping it to the Bateman-Feschbach-Tikochinsky oscillator and the Caldirola-Kanai oscillator. We have also discussed a couple of other black hole potentials to which this method can be applied.
%In this work we present a systematic procedure to quantize the black hole quasi-normal modes. 
%Using the field theoretic Lagrangian, we map the black hole quasi-normal modes  
%to the Bateman-Feschbach-Tikochinsky oscillator and Caldirola-Kanai 
%oscillator. We then discuss the conditions under which the black hole quasi-normal modes form 
%a complete set. Using the system-bath model, similar to the Bateman-Feschbach-Tikochinsky oscillator, 
%we show that a complete set of black hole quasi-normal modes can be quantized and obtain the general form of the generating 
%functional. Finally, we derive the asymptotic quasi–normal mode frequencies for a BTZ black hole and discuss the implications to toplogically massive gravity and new massive gravity.
\end{abstract}


\ccode{PACS numbers: 04.30.-w,04.62.+v,04.60.-m,04.70.-s,04.70.Dy}
%\maketitle
%body
%\balancecolsandclearpage
\section{Introduction}

Resonant modes play a central role in the phenomena of energy flow
among coupled systems as they provide characteristic information about
the physical system. In the case of gravitating systems, these
resonant modes are commonly referred as {\it quasi-normal modes}.
These are damped perturbations about a fixed background
(black holes or Neutron stars) that propagate to spatial infinity
\cite{Kokkotas,Berti:2009kk,2011-Konoplya.Zhidenko-RMP}.

To understand the importance of the black hole quasi-normal modes and why they 
have attracted attention over the last four decades, let us consider a 
non-spherical collapse of a gravitational object that results in a slightly perturbed 
black hole. The black hole will then reach it's
quiscent state by radiating away the perturbations in the form of
gravitational waves.  The radiation form the quasi-normal modes
spectrum of the black hole.  One can therefore define the quasi-normal
modes of a black hole as single-frequency oscillations of the black
hole spacetime that satisfy ingoing boundary conditons at the black
hole event horizon and outgoing boundary conditions at spatial
infinity \cite{Price-Husain}. The real part (that corresponds to the
frequency of the oscillation) and complex part (that corresponds to
the damping rate) of the frequencies are independent of the initial
perturbations and depend only on the properties of the black holes.
%Due to this property, over the last four decades, the black hole
%quasi-normal mode frequencies have attracted considerable amount of
%attention.

In the above way of describing the dynamics and the modes emanating to
infinity, quasi-normal modes are assumed to be purely classical and are 
not quantized. However, there have been indications that these
carry some information about quantum gravity
\cite{1999-KalyanaRama.Sathiapalan-MPLA,2000-Horowitz.Hubeny-PRD,1998-Hod-PRL,2003-Dreyer-PRL}.
Firstly, QNM has been shown to be an useful tool in understanding
AdS/CFT correspondence. In other words, it has been shown that there
is an one-to-one mapping of the damping time scales (evaluated via
simple QNM techniques) of black holes in Anti-de Sitter spacetimes and
the thermalization time scales of the corresponding conformal field
theory (which are, in general, difficult to compute)
\cite{Mann55,1999-KalyanaRama.Sathiapalan-MPLA,2000-Horowitz.Hubeny-PRD}. Secondly,
using Bohr's correspondence principle ``the transition frequencies at
high quantum numbers equate the classical oscillation frequencies'' it
was realised that one could possibly identify the transition frequency
between different black hole mass with the black hole quasi-normal
mode frequencies
\cite{1998-Hod-PRL,2003-Dreyer-PRL,Maggiore:2007nq}. This indeed has
provided the correct black hole entropy spectrum like other more
rigorous semiclassical quantization approaches \cite{Skakala:2013lza}.

It is important to note that although the quasi-normal frequencies are 
derived using quantum mechanical techniques, however, by definition these 
are purely classical (see, for instance, \cite{2011-Konoplya.Zhidenko-RMP,Berti:2009kk}). 
%However, it is not clear how reliable the values of these asymptotic frequencies. 
%In other words, if the quasi-normal modes are treated quantum mechanically will there 
%be a significant corrections to the asymptotic frequencies. In this work 
%we addres this issue. 
The question which we address in this work is the following: If these modes are 
indeed quantum mechanical (which do not preserve the total probability at later times) can they 
yield directly the quasi-normal mode frequencies and whether the asymptotic frequencies match 
with the classical calculations? Also, can the quantization procedure provide the link 
between the quasi-normal frequencies and the black hole entropy spectrum. 

To authors knowledge, the quantization of black hole quasi-normal modes was attempted 
by Kim \cite{Kim:2005ts}. Kim maps the black hole quasi-normal modes to  
a dissipative system, in particular, the Bateman-Feshbach-Tikochinsky (BFT) oscillator and showed that the two --- 
the black hole quasi-normal modes and the BFT oscillator --- systems have the same group 
structure and, hence, maps the quantum states of the BFT oscillator to that of the 
quasi-normal modes.

However, there are several drawbacks in Kim's approach: First, 
the BFT oscillator has two modes. One that decays in time and other that gets amplified 
in time. It is natural to map the decay modes to quasi-normal frequencies, however, as pointed
by Kim, it is not clear what is the role of the amplifying modes? Second, unlike the normal modes, 
the quasi-normal modes do not form a complete set of basis vectors in the sense that the 
system cannot be described as a sum over its QNMs, unless the black hole potential satisfy certain 
conditions \cite{Ching}. Kim's analysis does not 
look into this fundamental problem. Last, as is well-known, it is {\it only} possible to write down 
an equivalent Lagrangian that leads to the equation of motion for a dissipative harmonic oscillator 
\cite{1979-Ray-AJP,2002-Um.etal-PRep,2007-Gitman.Kupriyanov-EPJC,2011-Baldiotti.etal-PLA,Galley2013}. 
It is imperative to show that several of these explicitly time-dependent, equivalent Hamiltonians 
(Lagrangians) are related to each other by a canonical (point) transformations. 

In this work, we systematically address each of the above points and show that using Feynman and Vernon's 
approach one can quantize the quasi-normal modes. The Feynman-Vernon formulation involves integrating over the bath degrees of freedom. 
In the case of black holes, we show that naturally the bath degrees of freedom correspond to the ``stretched horizon''. The 
concept of the stretched horizon was originally treated by 't Hooft in his paper on the quantum nature of a black hole \cite{'tHooft1985727}. 
This description of the bath degrees of freedom of a black hole is consistent with other descriptions used in literature. For non-degenerate 
horizon, the stretched horizon description provides a universal framework and leads to finite results. 

%In Sec. (\ref{classical}) we address the third point above by 
%showing that the recent approach by Galley \cite{Galley2013}  can be used to map different equivalent 
%Lagrangians of the damped Harmonic oscillator. More importantly, we use the the field theoretic analogy to 
%obtain the mapping and address the first point. We give a physical interpretation for the doubling of the 
%variables when one includes dissipation to any physical system. 
Our main aim of this work is to obtain the quasi-normal mode frequencies from a completely quantum approach aka 
path integral quantization procedure. We start by introducing, in Sec. (\ref{completeness}), the conditions necessary for the completeness of these mode.
%In Sec. (\ref{completeness}), we discuss the second point 
%above by looking at the conditions necessary for the completeness of these modes and present these conditions.
%Our main aim of this work is to obtain the quasi-normal mode frequencies from a completely quantum approach aka 
%path integral quantization procedure.
To illustrate these conditions we discuss two examples in \ref{modelpot}, namely, the Regge-Wheeler potential and the rectangular potential well. As it is well-known the equation of motion containing the Regge-Wheeler potential, 
corresponding to the Schawrzschild black hole, can be mapped to Huen equation whose solutions are yet unknown 
\cite{2003-Ishkhanyan.Suominen-JPA,2007-Fiziev-JPCS}, we simplified the situation by modeling the 
Regge-Wheeler potential to Poschl-Teller potential whose solutions are known \cite{Ferrari-Mashhoon}. 
%We illustrate the conditions of completeness introduced in Sec. (\ref{completeness}) by considering a rectangular potential barrier 
%and obtain the quasi-normal modes frequencies for this potential.  
In \ref{path-int} we have explicitly shown the quantization procedure that we used to derive the quasi-normal modes of the $2+1$ dimensional 
BTZ black hole in Sec. (\ref{BTZ}). We have also discussed the oscillator models, in \ref{oscillators}, that initiated the current quantization procedure.  
%quantization procedure and we derive the quasi-normal frequencies. To give a more realistic example we consider the $2+1$ dimensional BTZ balck hole in Sec. (\ref{BTZ}). We systematically show how this quantization procedure can be used to derive the QNMs of the BTZ black hole. 
In Sec. (\ref{conclusion}) we discuss our result and present concluding remarks.

%In the appendix we have shown that a recent approach by Galley \cite{Galley2013}  can be used to map different equivalent 
%Lagrangians of the damped harmonic oscillator. More importantly, we use the the field theoretic analogy to 
%obtain the mapping and elucidate the concept of the amplifying modes. We give a physical interpretation for the doubling of the 
%variables when one includes dissipation to any physical system.

Throughout this work we use the Planck units $c=\hbar=G=k_{B}=1$.

\section{\label{completeness}Quantization of quasi-normal modes}

Before we go into the quantization, it may be useful to look at the
crucial differences between the normal and quasi-normal modes and the
issues involved in quantization of the quasi-normal modes: Let us
consider a system whose motion is defined by an equation of the
Sturm-Liouville type, for example a finite vibrating string, whose
motion is given by the wave equation. Since we can write the
Sturm-Liouville type equation as a self-adjoint operator and can use
the concept of normal modes to describe the dynamics of the system.
Each normal mode is an oscillation in which all the components of the
system move with the same frequency and the same phase
\cite{Goldstein}. The normal modes form a complete set and we can
write down the most general motion of the system as a superposition of
the normal modes \cite{Goldstein}.

\par{}Now suppose we introduce dissipation to this system, for example
we couple a semi-infinite string to the finite string via a spring
\cite{Schutz}. Such a system can dissipate energy away to infinity via
radiation. The dynamics of the system are described by quasi-normal
modes.  Quasi-normal modes, on the other hand, form complete sets only
under some special conditions \cite{Ching}. Therefore a system of
quasi-normal modes can be quantized only if those conditions are
satisfied. It should be noted that both canonical \cite{Leung-Suen, Ho} and path integral
\cite{Brink} quantizations of QNMs of leaky optical cavities have been
attempted.

In the rest of this section, we discuss the conditions required for the completeness 
of the quasi-normal modes and definition of the inner product and normalization of 
the quasi-normal modes.

\subsection{Completeness of QNMs} 
%As mentioned in Sec. (\ref{classical}) 
We use the Klein-Gordon (KG) equation to describe wave propagation in
curved space
\begin{align}
\label{Klein-Gordon}\tilde{D}\phi(x,t)\equiv\left[\partial_t^2 - \partial_x^2 + V(x)
\right]\phi(x,t) = 0.
\end{align}
%
%In this work, we focus on the Regge-Wheeler potential. 
%As shown in \cite{Chandra} it is possible to transform the above 
%equation for polar perturbation to Zerrilli potential for Axial 
%perturbations. Hence, we focus on the Regge-Wheeler 
%potential.
Assuming the time dependency of the form $e^{-i\omega t}$, the KG
equation reduces to a Schr\"{o}dinger-like equation
\begin{align}
\label{Schr}D\phi(x)\equiv\left[\partial_x^2+\omega^2-V(x)\right]\phi(x) = 0.
\end{align}
The QNMs are the eigenfunctions $f(x,t)$ of the operator $\tilde{D}$
or equivalently the eigenfunctions $f(x)$ of the operator $D$, that
satisfy the condition
\begin{align}
f(x)\sim e^{i\omega|x|},\quad |x|\to\infty.
\end{align}
If the potential $V(x)$ is finite everywhere and vanishes sufficiently
rapidly as $\mid x\mid\to\infty$ and there are spatial discontinuities
in $V(x)$ marking a \textquotedblleft cavity\textquotedblright $I$,
then for $x\in I$, $\lbrace f_j\rbrace$ is a complete set and we can
express the time evolution of the wavefunction as a sum over the QNMs,
$\phi(x,t) = \sum_ja_jf_j(x)e^{-i\omega_jt}$ or
$\phi(x,t=0)=\sum_ja_jf_j$ \cite{Ching, Price-Husain}.

\subsection{Inner product and normalization}

Let $\phi(x,t)$ and $\psi(x,t)$ be two wavefunctions in the space
spanned by $\{f_j(x,t)$ and let $\{f_j\}$ be complete. Therefore
$\phi(x,t=0)=\sum_ja_jf_j(x)$ and $\psi(x,t=0)=\sum_jb_jf_j(x)$.
Defining the two-component wavefunction
$\boldsymbol{\phi}=\left(\phi,\partial_t\phi\right)^T$ we give the
inner product between the two wavefunctions as \cite{Leung,
  PhysRevA.49.3057, Zhang}
\begin{align}\label{innerpro}
\left(\boldsymbol{\phi}|\boldsymbol{\psi}\right) = i\left\lbrace\int_{-a}^{a}dx\left[\phi(\partial_t\psi)+(\partial_t\phi)\psi\right]+\left[\phi(-a)\psi(-a)+\phi(a)\psi(a)\right]\right\rbrace
\end{align}
where $|a|$ is a finite value. The inner product gives the normalization relation between the QNMs
\begin{align}\label{orthonorm}
\left(\boldsymbol{f}_j|\boldsymbol{f}_j\right) = 2\omega_j\int_{-a}^{a}f_j^2(x)dx+i\left[f_j^2(-a)+f_j^2(a)\right]
\end{align}
where $\boldsymbol{f}_j = (f_j, -i\omega_jf_j)^T$ and the expansion
coefficients are given by
\begin{align}
a_{j}=\dfrac{\left(\boldsymbol{\phi}_m|\boldsymbol{f}_j\right)}{\left(\boldsymbol{f}_j|\boldsymbol{f}_j\right)},\quad
b_{j}=\dfrac{\left(\boldsymbol{\psi}_m|\boldsymbol{f}_j\right)}{\left(\boldsymbol{f}_j|\boldsymbol{f}_j\right)}.
\end{align}
Eqs. (\ref{innerpro}) and (\ref{orthonorm}) give the orthogonality
relation
\begin{align}\label{orthogon}
\int_{-a}^{a}f_j(x)f_k(x)dx = \delta_{jk}-i\left(\dfrac{f_j(-a)f_k(-a)+f_j(a)f_k(a)}{\omega_j+\omega_k}\right).
\end{align}

\section{\label{BTZ}Quasinormal modes of the BTZ black hole}
We consider a massless scalar field in a BTZ black hole background, whose metric is given by 
\begin{align}
ds^2=f(r)dt^2 - f(r)^{-1}dr^2-r^2d\varphi^2
\end{align} 
where $f(r) = -M+r^2+J^2/4r^2$. In the remainder of the article we consider the black hole mass, $M$, to be $1$ and the angular momentum, $J$, to be $0$. The external black hole horizon is at $r=r_H=M^{1/2}=1$ and the internal black hole horizon is at $r=0$. Spatial infinity is at $r\to\infty$. To make the following calculations easier we now introduce the tortoise coordinate, $dx=dr/f(r)$. In the tortoise coordinate, the horizon is mapped to $x\to-\infty$ and the spatial infinity to $x\to0$. The BTZ metric in the tortoise coordinate is given by
\begin{align}
ds^2 = f(r)[dt^2-dr^2] - r^2d\varphi^2
\end{align}
and the equation of motion of the massless scalar field is
\begin{align}
\left[\partial_t^2-\partial_x^2\right]\psi-\dfrac{f(r)}{r^2}\partial_\varphi^2\psi = 0.
\end{align}
Writing the scalar field as a sum of partial waves, $\psi=\sum_l\psi_l(t,x)\exp(il\varphi)$, we get
\begin{align}
\left[\partial_t^2-\partial_x^2\right]\psi_l + V_l(x)\psi_l=0
\end{align}
where the potential $V_l(x) = l^2f(r)/r^2$.

The BTZ black hole is asymptotically AdS. In the AdS geometry, fluctuations cannot escape to infinity and therefore dissipation of  energy by time-localized fluctuations happen at the black hole horizon \cite{Rodriguez}, i.e. the scalar field satisfies completely ingoing boundary conditions at the black hole event horizon. The dissipation can be considered to be due to interaction between the scalar field and a bath. As discussed in Sec. (\ref{completeness}) and \ref{modelpot}, the completeness of the QNMs requires us
to keep the potential finite. While this looks artificial, in the case of 
black hole, this corresponds to the fact that the region close to the 
horizon that leads to large blue-shifting of modes needs to be removed.
This naturally leads  to the concept of a stretched horizon, which, in turn, helps us to define a bath. We consider a stretched horizon, $r=r_H+\varepsilon = 1+\varepsilon$, and treat the variables between the stretched horizon and the horizon as the bath variables \cite{Iso}.  %Following the procedure in Sec. (\ref{path-int}) we get 
The equations of motion of the bath variables are
%we divide the region between $(-\infty,x(1+\varepsilon)]$ infinite segments as %$(x)_n = -N + nd$
%$x_n = x(1+\varepsilon) - nd$ (where $n = 0,1,2,\ldots,\infty$ and $d$ is the lattice spacing) and set harmonic oscillators $q_n$ at these lattice points. The equation of motion of the $j$-th oscillator is given by
%\begin{align}\label{joscill}
%\ddot{q}_j=-k(q_j-q_{j-1})+k(q_{j+1}-q_j)-V_l((x)_j)q_j
%\end{align}
%Using the forward and backward differentials
%$$
%\Delta^+q_j=\dfrac{q_{j+1}-q_n}{d},\;\;\;\Delta^-q_j=\dfrac{q_j-q_{j-1}}{d}
%$$
%and the relation 
%$$
%(q_{j-1}-q_j)-(q_j-q_{j+1}) = d\left(\Delta^+-\Delta^-\right)= d^2\Delta^+\Delta^-q_j
%$$
%we can write the equation of motion of the $j$-th oscillator as
%$$
%\ddot{q}_{j} = kd(\Delta^+ - \Delta^-)q_j-V_l((x)_j)q_j = kd^2\Delta^+\Delta^-q_j-V_l((x)_j)q_j
%$$
%which in the continuum limit ($d\to0$ and $kd^2=1$) gives
\begin{align}
\ddot{\psi}_l(t,x) = \partial_x^2\psi_l(t,x) - V_l(x)\psi_l(t,x). 
\end{align}
The value of the potential, $V_l(x)$ is vanishingly small near the horizon and can be neglected and the bath variables may be considered as free fields. Therefore the total action for the scalar field and the bath variables in Euclidean time is 
\begin{align} %\marginnote{how to include $S_{int}$}[0cm]
S &= \dfrac{1}{2}\int_{periodic} %_0^\beta d\tau\int_X 
d\mathbf{x}^2 \left[\left(\partial_\tau\psi_l\right)^2+\left(\partial_x\psi_l\right)^2+V_l(x)\psi_l^2\right]\\
& = S_{sys} + S_{bath}
\end{align}
where $S_{sys}$ is the action over the fields in $x\in(-\infty,0)$ and $S_{bath}$ is the action over the fields in the region $x\in(-\infty,x(1+\epsilon)]$ and $V_l(x) = 0,\;\text{for }x\in(-\infty,x(1+\varepsilon)]$.
%$$
%As in Sec. (\ref{path-int}) 
We consider the presence of a source $\chi=\sum_l\chi_l$ that interacts with the massless scalar field via 
\begin{align} %\marginnote{check the boundary terms}[0cm]
(\psi_l,\chi_l)=&i\left\{\int_{-\infty}^{0}dx\,\left[\psi_l(\partial_\tau{\chi_l})+(\partial_\tau{\psi_l})\chi_l\right]+\left[\psi_l(-\infty)\chi_l(-\infty) %+\psi_l(0)\chi_l(0)
\right]\right\}.
\end{align}
%then there will be an additional term, $(\psi_l,\chi_l)$, in $S_{sys}$.
The global generating functional is 
\begin{align}
W[\chi_l] = %\dfrac{1}{Z}
\int \mathcal{D}\psi_l\;e^{-S(\psi_l,\partial_\mu\psi_l)+(\psi_l,\chi_l)}.
\end{align}
Separating the global generating functional into system and bath factors and integrating out the latter, leaves us with
\begin{align}
W[\chi_l]^r = %\dfrac{1}{Z}
\int \mathcal{D}\psi_l^{(sys)}\;e^{-S_{sys}-S_{IF}+(\psi_l,\chi_l)}
\end{align} 
where the influence functional, $S_{IF}$ is defined by 
\begin{align}
e^{-S_{IF}} = & \int\mathcal{D}\psi_l^{(bath)}\;e^{-S_{bath}}.
\end{align}
To make the following calculations easier we approximate the bath action by assuming that the horizon is at $x=-N$ for some very large $N$ and the stretched horizon is at $x=-N+\delta$. The bath action under this approximation is
\begin{align}
S_{bath} & = \dfrac{1}{2}\int_0^\beta d\tau\int_{-N}^{-N+\delta}dx\left[(\partial_\tau\psi_l)^2 + (\partial_x\psi_l)^2\right] \nonumber\\
& = \dfrac{1}{2}\int_0^\beta d\tau\int_{0}^{\delta}d\xi\left[(\partial_\tau\psi_l)^2 + (\partial_x\psi_l)^2\right]
\end{align}
where $\xi = x+N$. Following the procedure in \ref{path-int}, we expand the bath fields,
\begin{align}
\psi_{l}^{(bath)} = T\sum_k\left\{\vphantom{\sum_{u=1}^\infty\psi_{l,k,u}\sin\left(\dfrac{\pi u\xi}{-\delta}\right)}\psi_{l,k}(-N)\dfrac{\delta-\xi}{\delta}+ \sum_{u=1}^\infty\psi_{l,k,u}\sin\left(\dfrac{\pi u\xi}{\delta}\right)\right\}e^{-i\nu_k\tau}.
\end{align}
As defined earlier, $T$ is the Hawking temperature and $\nu_k$ are the bosonic Matsubara frequencies. Therefore the influence functional is 
%\begin{widetext} 
%\vspace{-2em}
\begin{align}
e^{-S_{IF}} = &\exp\left[\dfrac{T}{2}\sum_k|\psi_{l,k}(-N)|^2\left[\dfrac{\delta\nu_k^2}{3}+\dfrac{1}{\delta}-\sum_{u=1}^\infty\dfrac{2(\delta\nu_k^2/\pi u)}{\delta\nu_k^2+\pi^2u^2/\delta}\right]\right]\\=&\exp\left[-\dfrac{T}{2}\sum_k|\psi_{l,k}(-N)|^2c(\delta,\nu_k)\right].
\end{align}
%\vspace{-1em}
%\end{widetext}
We expand the system fields as $\psi_l$ as $\psi_l = \sum_k\psi_{l,k}$ $= \sum_k \sum_ma_{lkm}f_{m}(x)\exp\{\omega_{m}\tau\}$ $ = \sum_k\sum_ma_{lkm}(\tau)f_{m}$. The modes $f_{m}(x)$ satisfy the ingoing boundary conditions at the horizon and if we impose the Dirichlet boundary conditions at $r\to\infty$, then these modes would correspond to the quasi-normal modes of the BTZ black hole. Similarly we expand the source, $\chi_l = \sum_k\sum_m b_{lkm}(\tau)f_{m}$. 
Now temporarily assuming that $V_l(x)=0,\forall\;x$, we calculate the effective generating functional for the system to be
%\begin{widetext} 
%\vskip-0.5in
%\vspace{-1em}
\begin{align} %\marginnote{to be simplified}
W[\chi_l]^r_0 &= \int\mathcal{D}\psi_l^{(sys)}\left[\exp\left\{T\sum_k
\left[- %2
\sum_{mn}a_{lkm}a_{lkn} \right.\right.\right.\nonumber\\&\times\left.\left.\left.\left( \omega_{m}\omega_{n}\delta_{mn}-  \dfrac{i\omega_{m}\omega_{n}f_{m}(-N)f_{m}(-N)}{\omega_{m}+\omega_{n}}\right)
+2\sum_m\omega_ma_{lkm}b_{lkm}\right]-S_{IF}\right\} %\left(\dfrac{\text{Det}\mathbf{M}}{\pi}\right)^{-1/2}
\right]
\end{align}
%\vspace{-1em}
%\end{widetext}
where %$\Delta a_{lkm}a_{lkn} = a_{lkm}(\beta)a_{lkn}(\beta) - a_{lkm}(0)a_{lkn}(0)$ and 
we have used the ``orthogonality'' condition $$
\int dx\; f_{m}(x)f_{n}(x) = \delta_{mn} - i\dfrac{f_{m}(-N)f_{n}(-N)}{\omega_{m}+\omega_{n}}.
$$
Before proceeding further we introduce the following expressions to simplify the remaining mathematical equations:
\begin{align*}
\mathcal{F}_{mn} &= \dfrac{f_m(-N)f_n(-N)}{\omega_m+\omega_n},\\\mathcal{I}&=\int_{-N}^{0} dx V_l(x)f_m(x)f_n(x),\\
\mathcal{O}_{kmn}&=\left(\theta(k)\dfrac{\omega_m}{\omega_m - ic(\delta,\nu_k)}+\theta(-k)\dfrac{\omega_n}{\omega_n+ic(\delta,\nu_k)}\right),\\
\mathcal{P}_{kmn}&=\left[1-iT\dfrac{\mathcal{F}_{mn}}{2\omega_m\omega_n}\mathcal{O}_{kmn}\left(1+iT\mathcal{F}_{mn}b_{lkm}b_{lkn}\right)\right].
\end{align*}
Now for the more general case of $\omega_m\neq\omega_n$, we can simplify the generating functional to
%\begin{widetext} 
%\vspace{-2em}
\begin{align}
W[\chi_l]^r_0 =& \int\mathcal{D}\psi_l^{(sys)}\exp\left[T\sum_k\left\{-\sum_{mn}a_{lkm}a_{lkn}f_m(-N)f_n(-N)\right.\right. \nonumber\\&\left.\left.\times\left(\dfrac{c(\delta,\nu_k)(\theta(k)\omega_m-\theta(-k)\omega_n)+i\omega_m\omega_n}{\omega_m+\omega_n}\right)+\sum_m 2\omega_ma_{lkm}b_{lkm}\right\}\right]\nonumber\\
=&\exp\left[iT\sum_{kmn}b_{lkm}b_{lkn}\mathcal{F}_{mn}\mathcal{O}_{kmn}\right].
%\vspace{-1em}
\end{align}
%\end{widetext}
Therefore the generating functional for the actual nonzero potential is
%\pagebreak
%\begin{widetext} \vspace{-2em}
\begin{align}\label{final_gen}
W[\chi_l]^r=&\exp\left[-\beta\int_{-N}^{0}dx\;V_l(x)f_m(x)f_n(x)\vphantom{\int_{-N}^{0}dx\;V_l(x)f_m(x)f_n(x)}\dfrac{1}{2\omega_m\omega_n}\dfrac{\partial}{\partial b_{lkm}}\dfrac{\partial}{\partial b_{lkn}}\right]W[\chi_l]^r_0.
\end{align}
%\vspace{-2em}\end{widetext}
The expression in the right-hand side of Eq. (\ref{final_gen}) can be calculated by perturbatively expanding the exponential. Taking only the first two terms of the perturbative expansion we get,
%\begin{widetext} 
%\vspace{-2em}
\begin{align}\label{complete-gen}
W[\chi_l]^r{=}&\exp\bigg\{iT\sum_{kmn} b_{lkm}b_{lkn}\mathcal{F}_{mn}\mathcal{O}_{kmn}\bigg\}\left[1-iI\dfrac{\mathcal{F}_{mn}\mathcal{O}_{kmn}}{2\omega_m\omega_n}\left(1+iT\mathcal{F}_{mn}b_{lkm}b_{lkn}\right)\right].
\end{align}
%\begin{align}\label{complete-gen}
%W[\chi_l]^r {=} &\exp\bigg\{ iT\sum_{kmn}b_{lkm}b_{lkn}\dfrac{f_m(-N)f_n(-N)}{\omega_m+\omega_n}\left(\theta(k)\dfrac{\omega_m}{\omega_m - ic(\delta,\nu_k)}+\theta(-k)\dfrac{\omega_n}{\omega_n+ic(\delta,\nu_k)}\right)\bigg\}\nonumber\\
%& \times \left[1 - i\int_{-N}^{0} dx\;V_l(x)f_m(x)f_n(x)\dfrac{f_m(-N)f_n(-N)}{2\omega_m\omega_n(\omega_m+\omega_n)}\right.\nonumber\\ & \left. \times\left(\theta(k)\dfrac{\omega_m}{\omega_m - ic(\delta,\nu_k)}+\theta(-k)\dfrac{\omega_n}{\omega_n+ic(\delta,\nu_k)}\right)\left[1+iT\dfrac{f_m(-N)f_n(-N)}{\omega_m+\omega_n}b_{lkm}b_{lkn}\right]\vphantom{V_l(x)f_m(x)f_n(x)\dfrac{f_m(-N)f_n(-N)}{2\omega_m\omega_n(\omega_m+\omega_n)}\left(\theta(k)\dfrac{\omega_m}{\omega_m - ic(\delta,\nu_k)}+\theta(-k)\dfrac{\omega_n}{\omega_n+ic(\delta,\nu_k)}\right)}\right].
%\end{align}
%\vspace{-1em}
%\end{widetext}
The generating functional in Eq. (\ref{final_gen}) is related to the temperature Green's function via
\begin{align}
\left.\dfrac{\partial^2W[\chi_l]^r}{\partial b_{lkm}\partial b_{lkn}}\right|_{\chi = 0} = -4\omega_m\omega_n\mathcal{G}_{mn}(\nu_k).
\end{align}
Double differentiating Eq. (\ref{complete-gen}) we get,
%\begin{widetext} 
%\vspace{-2em}
\begin{align}\label{double_diff}
\dfrac{\partial^2 W[\chi_l]^r}{\partial b_{lkm}\partial b_{lkn}}=&\exp\bigg\{iT\sum_{kmn} b_{lkm}b_{lkn}\mathcal{F}_{mn}\mathcal{O}_{kmn}\bigg\}\left[\vphantom{\dfrac{\mathcal{F}_{mn}^3}{2\omega_m\omega_n}}\left(iTb_{lkm}\mathcal{F}_{mn}\mathcal{O}_{kmn}\right)^2\mathcal{P}_{kmn}\right.\nonumber\\&\qquad{}\left.+iT\mathcal{F}_{mn}\mathcal{O}_{kmn}\mathcal{P}_{kmn} +i\mathcal{I}T^2b_{lkm}b_{lkn}\dfrac{\mathcal{F}_{mn}^3\mathcal{O}_{kmn}^2}{2\omega_m\omega_n}\right.\nonumber\\&\qquad{}\left.+i\mathcal{I}T^2\dfrac{\mathcal{F}_{mn}^3\mathcal{O}^2_{kmn}}{2\omega_m\omega_n}b_{lkn}^2+\mathcal{I}T\dfrac{\mathcal{F}_{mn}^2\mathcal{O}_{kmn}}{2\omega_m\omega_n}\right].
\end{align}
%\vspace{-0.5em}
%\end{widetext}
The condition that the source, $\chi=0$, implies that the coefficients $b_{klj}$ are all zero. Applying this condition to Eq. (\ref{double_diff}) we get
%%\begin{widetext}
\begin{align}
\left.\dfrac{\partial^2W[\chi_l]^r}{\partial b_{lkm}\partial b_{lkn}}\right|_{\chi = 0}= iT\mathcal{F}_{mn}\mathcal{O}_{kmn}+\dfrac{T^2\mathcal{F}^2_{mn}\mathcal{O}^2_{kmn}}{2\omega_m\omega_n}+\dfrac{\mathcal{I}T\mathcal{F}_{mn}^2\mathcal{O}_{kmn}}{2\omega_m\omega_n}.
\end{align}
%\end{widetext}
The temperature Green's function therefore turns out to be
%\begin{align}
%	\mathcal{G}_{mn}(\nu_k)=&-\dfrac{1}{4\omega_m\omega_n}\left[iT\mathcal{F}_{mn}\mathcal{O}_{kmn}+\dfrac{T^2\mathcal{F}^2_{mn}\mathcal{O}^2_{kmn}}{2\omega_m\omega_n}\right.\nonumber\\&\qquad{}\left.+\dfrac{\mathcal{I}T\mathcal{F}_{mn}^2\mathcal{O}_{kmn}}{2\omega_m\omega_n}\right]
%\end{align}
%which, for the ease of calculation, we expand to
%\begin{widetext} 
%\vspace{-2em}
\begin{align}
%\mathcal{G}_{mn}(\nu_k)=&-\dfrac{1}{4\omega_m\omega_n}\left[iT\mathcal{F}_{mn}\mathcal{O}_{kmn}+\dfrac{T^2\mathcal{F}^2_{mn}\mathcal{O}^2_{kmn}}{2\omega_m\omega_n}+\dfrac{\mathcal{I}T\mathcal{F}_{mn}^2\mathcal{O}_{kmn}}{2\omega_m\omega_n}\right]
%\\\begin{align}
\mathcal{G}_{mn}(\nu_k)&=-\dfrac{1}{4\omega_m\omega_n}\left[iT\mathcal{F}_{mn}\mathcal{O}_{kmn}+\dfrac{T^2\mathcal{F}^2_{mn}\mathcal{O}^2_{kmn}}{2\omega_m\omega_n}+\dfrac{\mathcal{I}T\mathcal{F}_{mn}^2\mathcal{O}_{kmn}}{2\omega_m\omega_n}\right]\\
\label{temp-green}
&=\dfrac{T\mathcal{F}_{mn}}{8\omega_m^2\omega_n^2}\dfrac{\left[\omega_m\omega_n+ic(\delta,\nu_k)\left(\theta(k)\omega_m-\theta(-k)\omega_n\right)\right]}{\left(\omega_m-ic(\delta,\nu_k)\right)\left(\omega_n+ic(\delta,\nu_k)\right)}\nonumber\\
&{}\times\left[\left(i2\omega_m\omega_n+\mathcal{I}T\mathcal{F}_{mn}\right)+\dfrac{T\mathcal{F}_{mn}\left(\omega_m\omega_n+ic(\delta,\nu_k)\left(\theta(k)\omega_m-\theta(-k)\omega_n\right)\right)}{\left(\omega_m-ic(\delta,\nu_k)\right)\left(\omega_n+ic(\delta,\nu_k)\right)}\right].
\end{align}
%\vspace{-0.5em}
%\end{widetext}
The temperature Green's function is related to the real-time retarded propagator as 
$
\mathcal{G}_{mn}(\nu_k)=G^R_{mn}(i\nu_k)
$
i.e., we make the substitution $c(\delta,\nu_k)\to c(\delta,i\nu_k)$ in Eq. (\ref{temp-green}) to get the real time retarded propagator. We can then compute the QNM frequencies as the poles of $G^R_{mn}$, i.e.,
%\begin{align}
$\omega_n = 0\;\text{or}\;\omega_n = \pm ic(\delta,i\nu_k)$.
%\end{align}
We can express the function $c(\delta,i\nu_k)$ in terms of the digamma function, which is the logarithmic derivative of the gamma function,
\begin{align}\label{c-digamma}
c(\delta,i\nu_k) = \alpha_k-\dfrac{1}{\pi}\left(\psi^{(0)}(1-s_k)+\psi^{(0)}(1+s_k)\right)
\end{align}
where $\psi^{(0)}$ is the digamma function,
\begin{align}
s_k = \frac{\delta \, \nu_k}{\pi}; \quad \alpha_k=- \frac{\delta \, \nu_k^2}{3} + \frac{1}{\delta}- \frac{2 \, \gamma}{\pi}
\end{align}
and $\gamma$ is the Euler-Mascheroni coefficient. The digamma function can be expressed as a converging rational zeta series if $|s_k|<1$. We can ensure that this condition is satisfied by choosing a sufficiently small $\delta$. The expression $\psi^{(0)}(1-s_k)+\psi^{(0)}(1+s_k)$ tends to infinity as $|s_k|\to1$. Therefore to get the asymptotic QNM frequencies we expand $\psi^{(0)}(1-s_k)+\psi^{(0)}(1+s_k)$ at $s_k=1$,
\begin{align}
\psi^{(0)}(1- s_k)+\psi^{(0)}(1+s_k)=  \dfrac{1}{1-s_k}+(1-2\gamma)+\dfrac{1}{3}(\pi^{2}-3)(s_k-1) + \mathcal{O}(s_k^{2}).
\end{align}
%By making the value of $\delta$ sufficiently small we can ensure that $|s|<1$ and thereby expand the digamma function in a converging rational zeta series
%\begin{align}\label{zeta}
%\psi^{(0)}(1+s)=-\gamma-\sum\limits_{k=1}^{\infty}\zeta(k+1)(-s)^k
%\end{align}
%where $\zeta$ is the Riemann zeta function. Substituting Eq. (\ref{zeta}) in Eq. (\ref{c-digamma}) we get
%\begin{align}
%c(\delta,i\nu_k) = \dfrac{T}{2}\left[\alpha+\dfrac{1}{\pi}\sum_{j=1}^{\infty}\zeta(j+1)\left\lbrace(-s)^j+(s)^j\right\rbrace\right]
%\end{align}
%with $\alpha$ being redefined as $1/\delta-\delta\nu_k^2/3$. 
%Thus the QNM frequencies are 
%\begin{align}\label{qnf-last-but-one}
%\omega_n = \pm \dfrac{iT}{2}\left[\alpha+\dfrac{1}{\pi}\sum_{j=1}^{\infty}\zeta(j+1)\left\lbrace(-s)^j+(s)^j\right\rbrace\right].
%\end{align}
%The expression $\psi^{(0)}(1-s)+\psi^{(0)}(1+s)$ tends to infinity as $|s|\to1$. Therefore to get the asymptotic QNM frequencies we apply the limit $|s|\to1$ on Eq. (\ref{qnf-last-but-one}), obtaining
%\begin{align}
%\omega_n^{\text{asym}} =  \pm\dfrac{iT}{2}\left[\alpha+\dfrac{1}{\pi}\sum_{j=1}^{\infty}\zeta(j+1)\left\lbrace(-1)^j+(1)^j\right\rbrace\right]
%\end{align}
%which simplifies to 
%\begin{align}
%\omega_n^{\text{asym}} = \pm\dfrac{iT}{2}\left(\alpha-\dfrac{1}{\pi}\right).
%\end{align}
%The Eq. (\ref{c-digamma}) can be simplified by using the identity $$\psi^{(0)}(1-s)+\psi^{(0)}(1+s) = H_{-s} + H_{s} + 2\gamma$$.{}
The asymptotic QNM frequencies are therefore
\begin{align}\label{finalfreq}
\omega_{n}^{\text{asym}} =  \pm i\left\lbrace\alpha_k+\dfrac{1}{\pi}\left(\dfrac{1}{1-s_k}+(1-2\gamma)+\dfrac{1}{3}(\pi^{2}-3)(s_k-1) \right)\right\rbrace.
\end{align}
As we have mentioned we have to choose a sufficiently small $\delta$ to derive the asymptotic QNM frequencies, one obvious point of curiosity would be the nature of the frequencies in the limit $\delta\to0$. However this limit doesn't exist for the quantities $\alpha_k$ and $s_k$. To overcome this, we can consider $\delta$ to be of the order of $k^{-1}$ for very large $k$ (as defined in \ref{path-int}, $k\in\mathbb{Z}$). Therefore under the limit $\delta\to1/k$ the asymptotic QNM frequencies turn out to be
\begin{align}\label{freq_large_k}
\omega_n^{\text{asym}}\sim\pm\dfrac{iT}{2}k.
\end{align}

The BTZ black hole metric had originally been obtained as a solution to the $(2+1)$-dimensional version of classical Einstein-Hilbert theory of gravitation with a negative cosmological constant. However this version of the Einstein-Hilbert theory does not allow propagating bulk degrees of freedom. To resolve this problem, alternative $(2+1)$-dimensional theories of gravitation, such as the topologically massive gravity (TMG) and new massive gravity (NMG), have been proposed. Both TMG and NMG admit the BTZ metric as a solution. In recent years, there have been numerous studies concerning the QNMs of BTZ black holes in all the three theories of gravity; many of which have dealt with massive scalar fields in more generalized BTZ backgrounds.
All these classical calculations \cite{Govindarajan-Suneeta,Cardoso-Lemos,Birmingham-Choptuik,Sachs-Solodukhin} indicate that the scalar QNM frequencies of the BTZ black hole are of the form 
$$
\omega = l-i\mu T(k+1)
$$
$(l\;\text{is the angular momentum of the scalar field and }\mu\;\text{is a constant})$ which in the limit of large $k$ becomes
$$
\omega\sim -iTk.
$$
This result is similar to our calculations. More general calculations \cite{Kwon-Nam-Park-Yi}, in the context of NMG, show that QNM frequencies are of the form $$\omega=-iT\dfrac{h(k^2)}{g(k)}$$
where $h$ and $g$ are functions of $k$, and in the asymptotic limit becomes\begin{align}\omega\sim-i\dfrac{T}{2}k.\end{align} 
Though we have worked with a massless scalar field, the frequencies given by Eq. (\ref{finalfreq}) are similar in structure to those obtained in the case of NMG and coincide in the limit of large $k$. Moreover, from Eq. (\ref{freq_large_k}), we can see that the QNM frequencies are proportional to the Hawking temperature, just like in the case of the classical models. 

From the point of view of NMG, black holes with QNM frequencies of the form, $\omega = \omega_R+i\omega_I,\;\omega_R=0\;\text{and}\;\omega_I>0$, are unstable \cite{Myung-Kim-Moon-Park}. This is because modes with positive $\omega_I$ will grow exponentially with time. Considering the classical QNM spectra, the BTZ black hole are stable in NMG. However, the quantum QNM spectra, that we obtained, has modes with either positive or negative imaginary parts and therefore doesn't confirm to this stability condition. One way to introduce stability would be to consider a finite time frame so that QNMs whose frequencies have positive imaginary part do not blow up. Another possible way would be consider a theory of (2+1)-dimensional gravity that naturally introduces a cutoff for the QNM frequencies. To authors knowledge, none of the competing classical theories of gravity do that.

\section{\label{conclusion}Conclusion}
In this paper we have presented two descriptions of the QNMs - one
classical and the other quantum mechanical. In the classical picture (\ref{oscillators})
we have drawn analogy between the QNMs and the damped simple harmonic
oscillator. We have shown that there's a connection between the system
of QNMs and the Bateman oscillator and therefore the study of one
system may give valuable information about the other. The damped
harmonic oscillator is considered to be coupled to an amplified one
and the system therefore follows the Bateman equations of motion.  

We used the path integral formalism prescribed by Feynman and Vernon to develop our quantum theory. This is because the system-bath model of the Feynman-Vernon formalism is in spirit similar to Bateman's dual oscillator system and the physics can therefore be intuitively understood. Like in the case of the Bateman dual oscillator, the quantization of the system-bath model involves integrating over the bath degrees of freedom. We have given a consistent description of the bath degrees of freedom for a black hole using the concept of a stretched horizon, originally developed by 't Hooft.

In Sec. (\ref{BTZ}) we have given a detailed derivation of the QNM frequencies of a BTZ black hole, using our quantum theory. The asymptotic QNM frequencies in this case matches with the classical results. However, our results do not 
show that the imaginary part of the asymptotic QNM frequencies are quantized.

Moreover the final forms of the generating functional obtained in Sec. (\ref{BTZ}) and \ref{path-int} can be used to derive interesting physical quantities, such as the free energy, defined as, $F_{c}^{R}[\chi] \equiv \ln W_c^{R}[\chi]$ and entropy. We hope to present these applications of the generating functional in a future article.

\section*{Acknowledgments}
The authors thank Sashideep Gutti for discussions. SP thanks IISER-TVM for 
hospitality for carrying out the initial part of the work and is supported 
by Junior Research Fellowship of CSIR, India. KR is supported by the INSPIRE fellowship, DST, India. The work is supported by Max Planck-India Partner Group on Gravity and Cosmology. SS is partially supported by Ramanujan Fellowship of DST, India.

\appendix 
\section{\label{modelpot}Model Potentials}

Taking into account the restrictions set on the potential in Sec.
(\ref{completeness}), we will use potentials that have support in the
interval, $I\equiv[-a,a]$. In this section we will describe two such
potentials and the associated QNMs.

The rectangular barrier is given below as an illustration of the method 
as it gives exact quasi-normal mode frequencies while the modified 
P\"{o}schl-Teller potential has all the features of Regge-Wheeler potential 
\cite{Ferrari-Mashhoon}.

\subsection{Rectangular barrier}
The rectangular barrier potential is given by
\begin{align}\label{barrier}
V(x)=\begin{cases}
V_0\neq0,\; x\in[-a,a]\\
0,\; \text{otherwise}.
\end{cases}
\end{align} 
For this barrier potential the boundary conditions that should be satisfied by the QNMs are reduced to 
$$\dfrac{\partial_x f(x)}{f(x)} = \pm i\omega,\quad x=\pm a$$
The eigenfunctions of $D$ are given by Eq. (\ref{barrier})
\begin{align}
f(x)=\begin{cases}
Pe^{i\omega x},\; x>a\\
Qe^{ikx} + R^{-ikx},\; a\geq x\geq -a\\
Se^{-i\omega x},\; -a>x,
\end{cases} 
\end{align}
where $k=\sqrt{\omega^2 - V_0}$ and $P$, $Q$, $R$ and $S$ are constants. Maintaining the continuity of $f(x)$ and $\partial_xf(x)$ we obtain 
$$ \left(\dfrac{k-\omega}{k+\omega}\right)^2e^{i4ka} = 1 $$ which is simplified to
\begin{align}\label{kv0w}
k=i\omega_0\cos(ka)\quad\text{or}\quad k=-i\omega_0\sin(ka),
\end{align}
where $\omega_0 = V_0^{1/2}$. Depending on the sign of $V_0$, Eq.
(\ref{kv0w}) gives different sets of QNM frequencies \cite{Boonserm}.

\subsubsection{Negative $V_0$}
For $V_0<0$, $\omega_0$ is purely imaginary, $\omega_0=i|\omega_0|$.
Therefore from Eq. (\ref{kv0w})
\begin{align}\label{kor1}
k = -|\omega_0|\cos(ka)\quad\text{or}\quad k=|\omega_0|\sin(ka).
\end{align}
Eq. (\ref{kor1}) sometimes have solutions for real $k$ if
$|k|\leq|\omega_0|$. Since $\omega^2 = k^2 + \omega_0^2 = k^2 -
|\omega_0|^2\leq0$, therefore the values of $\omega$ are purely 
imaginary corresponding to bound states of the potential and not 
quasi-normal modes.\\[2mm]
However, if $\Im(k)\gg0$, then $\cos(ka)\approx\exp(-iak)/2$ and
$\sin(ka)\approx-\exp(-iak)/2$. In such a situation the approximate
values of the quasi-normal frequencies are given by the $j$-th
solutions of
\begin{align}
2k\approx-|\omega_0|e^{-ika}.
\end{align} 

\subsubsection{Positive $V_0$}
For $V_0>0$, $\omega_0$ is real. If $\omega = i|\omega|$,
corresponding to a purely damped mode, then from Eq. (\ref{kv0w})
\begin{align}
|k| = \omega_0\cosh(|k|a)\quad\text{or}\quad k=0.
\end{align}
The $k=0$ mode is not a true quasi-normal frequency, the other equation
gives the physical quasi-normal frequencies
\begin{align}
|k|a=\omega_0a\cosh(|k|a).
\end{align}
For small $\omega_0a$ there are two quasi-normal frequencies, one of
which is given by the perturbative expression
\begin{align}
k &= i\omega_0\left\lbrace 1+\dfrac{1}{2}(\omega_0a)^2+\dfrac{13}{24}(\omega_0)^4+\mathcal{O}([\omega_0a]^6)\right\rbrace\\
\text{or}\quad
\omega &= i\omega_0(\omega_0a)\left\lbrace 1+\dfrac{2}{3}(\omega_0a)^2+\dfrac{4}{5}(\omega_0a)^4+\mathcal{O}([\omega_0a]^6)\right\rbrace.
\end{align}
No such representations exist for the other quasi-normal mode. However if 
$\omega_0a \sim 0.663$, then the two quasi-normal modes merge and for 
$\omega_0a \gtrsim 0.663$ there are no quasi-normal modes.\\[2mm]
In this case, if $\Im(k)\gg0$ then the quasi-normal frequencies are the
$j$-th solutions of
\begin{align}
2k\approx-i\omega_0e^{-ika}.
\end{align}

\subsection{Modified P\"{o}schl-Teller potential}

The P\"{o}schl-Teller potential is a continuous function over the real
line $(-\infty,\infty)$. The QNMs of this potential forms a complete
set, in the sense that the solutions at late times can be represented
by an infinite sum of the QNMs \cite{Beyer}. However the quantization
procedure that we have used requires that
spatial discontinuities be present in the potential. We therefore
artificially introduce such discontinuities in the potential by making
it go to zero for $x > a$ for a large $a$ where the value of the
potential is very small\cite{Ching}. Therefore the modified
P\"{o}schl-Teller potential is
\begin{align}\label{modpt}
V(x)=\begin{cases}
\dfrac{1}{4\nu}(\nu^2+c^2)\sech^2(\sqrt{\nu} x),\; x\in[-a,a]\\
0,\; \text{otherwise}.
\end{cases}
\end{align} 
where the factors $\nu$ and $c$ can be adjusted to approximate actual
black hole potentials \cite{Ferrari-Mashhoon}, subject to the
condition $(\nu^2+c^2) \, > \, 0$. Here we have considered the quasi-exactly solvable 
form of the Poschl-Teller potential \cite{ChoQES,Ozer-Roy}, a special form of the
more general Scarf II potential.  The Schr\"{o}dinger - like equation
for the P\"{o}schl-Teller potential is
\begin{align}\label{SchroPT1}
\dfrac{d^2f(x)}{dx^2} + \omega^2f(x)=\begin{cases}
\dfrac{1}{4\nu}(\nu^2+c^2)\sech^2(\sqrt{\nu} x)f(x),\quad x\in[-a,a]\\
0,\quad\text{otherwise}.
\end{cases} 
\end{align}
We can approximate the QNM boundary condition as 
$$
f(x) = \left(\cosh(\sqrt{\nu}x)\right)^{(\nu+ic)/2\nu},\;\;x=\pm a.
$$ 
With these boundary conditions, we rewrite the wavefunction as 
$f = \left(\cosh(\sqrt{\nu}x)\right)g(x)$ \cite{Ozer-Roy}. Introducing the
new variable $y=\sinh(\sqrt{\nu}x)$ Eq. (\ref{SchroPT1}) becomes
\begin{align}\label{SchroPT1a}
\dfrac{d^2g(y)}{dy^2}+\dfrac{iy(c-2i\nu)}{(y^2+1)\nu} \dfrac{dg(y)}{dy}-\dfrac{c^2-2i\nu c-\nu(4\omega^2+\nu)}{4(y^2+1)\nu^2}g(y)=0.
\end{align}
%With the boundary conditions in mind we rewrite the wavefunction as $f(y)=(1-y)^{-i\omega/2}(1+y)^{-i\omega/2}g(y)$ and Eq. \ref{SchroPT1a} is reduced to
%\begin{align}\label{SchroPT1b}
%\dfrac{d^2g}{dy^2}=\dfrac{2y(1-i\omega)}{1-y^2}\dfrac{dg}{dy}+\dfrac{1-2i\omega-2\omega^2}{2(1-y^2)}g
%\end{align}
From Eq. (\ref{SchroPT1a}), using the asymptotic iteration method we get the QNM frequencies as \cite{Ozer-Roy,Cho:2011sf}
\begin{align}
\omega_n = \pm\left(\dfrac{c}{2\sqrt{\nu}}-i\dfrac{(2n+1)\sqrt{\nu}}{2}\right)
\end{align}
and the QNMs as
\begin{equation*}
f_n(x) \approx \left(\cosh(\sqrt{\nu}x)\right)^{\nu+ic/2\nu} P_n^{(\alpha,\beta)}\left(i\sinh(\sqrt{\nu}x)\right)
\end{equation*}
where $P_n^{(\alpha,\beta)}(x)$ are the Jacobi Polynomials.
%The QNMs for this potential are
%%
%\begin{align}
%f(x) = \begin{cases}
%Ae^{i\omega x},\; x>a\\
%B(1-\xi^2)^{-ik/2\alpha}F[w,x,y,z],\; a\geq x\geq -a\\
%Ce^{-i\omega x},\; -a>x
%\end{cases}
%\end{align}
%%
%where $k=\sqrt{\omega^2-V(x)}$, $F[w,x,y,z] =
%F[-ik/\alpha-s,-ik/\alpha+s+1,1-ik/\alpha,(1-\xi)/2]$, $\xi =
%\tanh(\alpha x)$, and $s=(-1+\sqrt{1+4V_0/\alpha^2})/2$ and $A$, $B$
%and $C$ are constants\cite{Landau, Ferrari-Mashhoon}.
%The continuity of $f(x)$ and $\partial_xf(x)$ across the borders at $x=\pm a$ gives us the transcendental equation for the QNM frequencies
\section{\label{path-int}Path Integral Quantization}

Classical treatments of non-conservative or open system often use the
cavity-bath model to describe such systems. The system of interest
forms the cavity and the surroundings forms the bath with infinite
degrees of freedom. Energy is exchanged between the cavity and the
bath via some interaction. The complete system is described by a
Lagrangian of the form $L(q,\dot{q},Q,\dot{Q}) =
L_c(q,\dot{q})+L_b(Q,\dot{Q})+L_{int}(q,\dot{q},Q,\dot{Q})$, where
$L_c(q,\dot{q})$ describes the cavity, $L_b(Q,\dot{Q})$ the bath and
$L_{int}(q,\dot{q},Q,\dot{Q})$ gives the interaction between the two
\cite{Caldeira-Leggett}. The structure of the Lagrangian shows that
the quantization would involve both the cavity $(q,\dot{q})$ and the
bath $(Q,\dot{Q})$ degrees of freedom. However we are not
interested in the time evolution of the bath. It is therefore
desirable to eliminate $Q$ and $\dot{Q}$ variables and compute the
expectation value of any observable in terms of $q$ and $\dot{q}$
only. Feynman and Vernon showed that the path integral formulation is
very effective in this regard \cite{Feynman}. The general idea is to
write down the generating functional or the density matrix of the
whole system and then integrate out the bath degrees of freedom. This
leaves us a reduced density matrix which incorporates the effect of
the bath on the cavity and is expressed in terms of the $q$ and
$\dot{q}$ variables only \cite{Feynman, weiss1999quantum}. In the case of a black hole these bath degrees of freedom can be described as the fields populating the space between the black hole event horizon and a stretched horizon.

\par{}As mentioned in Sec. (\ref{completeness}), QNMs are the
exponentially damped eigensolutions of the operator $\tilde{D}$ and
form a complete set for the potentials with discontinuities, like those mentioned in
Sec. (\ref{modelpot}). Moreover the spatial discontinuities in potentials allow us to treat the space around the source of the
QNMs as an open system or cavity coupled to a bath, the interval $I$
being the cavity and everything outside $I$ forms the bath; the
gravitational waves carry off energy from the cavity to the bath. The
QNMs can therefore be used for exact eigenfunction expansions in the
cavity which is analogous to the normal mode expansions in Hermitian
systems. If we eliminate the bath modes then we can second quantize
the open system in terms of the damped QNMs only. In this section we
will use the Feynman-Vernon formalism as employed in \cite{Brink} to
quantize QNMs of the model potentials.

\subsection{Integrating out the bath modes}
To start with, we consider the universe to be restricted to
$[-\lambda,\lambda]$ %We will later take $\lambda\to\infty$. 
and the cavity to $[-a,a]$.
The region $[-\lambda,-a] \cup [a,\lambda]$ forms the thermal bath. To get the equation of motion of the bath variables we divide the region between $[-\lambda,-a]$ into infinite segments as %$(x)_n = -N + nd$
$x_n = x(1+\varepsilon) - nd$ (where $n = 0,1,2,\ldots,\infty$ and $d$ is the lattice spacing) and set harmonic oscillators $q_n$ at these lattice points. The equation of motion of the $j$-th oscillator is given by
\begin{align}\label{joscill}
\ddot{q}_j=-\kappa(q_j-q_{j-1})+\kappa(q_{j+1}-q_j)
\end{align}
Using the forward and backward differentials
$$
\Delta^+q_j=\dfrac{q_{j+1}-q_n}{d},\;\;\;\Delta^-q_j=\dfrac{q_j-q_{j-1}}{d}
$$
and the relation 
$$
(q_{j-1}-q_j)-(q_j-q_{j+1}) = d\left(\Delta^+-\Delta^-\right)= d^2\Delta^+\Delta^-q_j
$$
we can write the equation of motion of the $j$-th oscillator as
$$
\ddot{q}_{j} = \kappa d(\Delta^+ - \Delta^-)q_j = \kappa d^2\Delta^+\Delta^-q_j
$$
which in the continuum limit ($d\to0$ and $kd^2=1$) gives
\begin{align}
\ddot{\phi}(t,x) = \partial_x^2\phi(t,x). 
\end{align} %is an open system, the entire universe is closed
%and is defined by the Lagrangian
%
The bath variables can therefore be considered as free fields. With this definition of the bath variables we can describe the universe by the Euclidean action
\begin{align} %\marginnote{how to include $S_{int}$}[0cm]
S &= \dfrac{1}{2}\int_{periodic} %_0^\beta d\tau\int_X 
d\mathbf{x}^2 \left[\left(\partial_\tau\phi\right)^2+\left(\partial_x\phi\right)^2+V(x)\phi^2\right]\\
& = S_{sys} + S_{bath}
\end{align}
%\begin{align*}
%L = \int_X dx\mathcal{L} = \dfrac{1}{2}\int_X dx \left[\left(\partial_t\phi\right)^2-\left(\partial_x\phi\right)^2-V(x)\phi^2\right]
%\end{align*}
%
where $V(x)=0,\;\text{for}\; x\in[-\lambda,-a] \cup [a,\lambda]$. %Introducing Euclidean time, $\tau=it$, we rewrite the Lagrangian,
%%
%\begin{align}
%L_E= -\dfrac{1}{2}\int_X dx \left[\left(\partial_\tau\phi\right)^2+\left(\partial_x\phi\right)^2+V(x)\phi^2\right].
%\end{align}
%%
%The Euclidean action for the universe is
%%
%\begin{align}
%S_E= -\dfrac{1}{2}\int_{0}^{\beta}d\tau\int_X dx \left[\left(\partial_\tau\phi\right)^2+\left(\partial_x\phi\right)^2+V(x)\phi^2\right].
%\end{align}
%%
%where $\beta$ is the inverse temperature. 
We also consider a source $\chi$
that interacts only with the cavity fields via
%\begin{widetext} \vspace{-2em}
\begin{align}
(\phi,\chi)=i\left\{\int_{a}^{-a}dx\left[\phi(\partial_\tau{\chi})+(\partial_\tau{\phi})\chi\right]+\left[\phi(-a)\chi(-a)+\phi(a)\chi(a)\right]\right\}.
\end{align}
%\vspace{-2em}\end{widetext}
%
Therefore the configuration space generating functional of the
cavity-bath system is
\begin{align}
W[\chi] = \dfrac{1}{Z}\int \mathcal{D}\phi\;e^{-S(\phi,\partial_\mu\phi)+(\phi,\chi)}.
\end{align}
The real field $\phi$ satisfies the boundary conditions
$\phi(-\lambda-a,0)=\phi(a+\lambda,0)=0$ and
$\phi(x,0)=\phi(x,\beta)$, with $\beta$ being the inverse of the Hawking temperature, $T^{-1}$. We now separate the generating functional
into cavity and bath factors, $Z^{-1}\int
\mathcal{D}\phi=Z_c^{-1}\int\mathcal{D}\phi_c\times
Z_b^{-1}\int\mathcal{D}\phi_b$, the latter running over the fields in
the region $[-\lambda-a,-a]\cup [a,a+\lambda]$, with given boundary
values $\phi(a,\tau)$ and $\phi(-a,\tau)$. $Z^{-1}$, $Z_{b}^{-1}$ and $Z_{c}^{-1}$ are normalizing factors. The bath factor turns out to be
%\begin{widetext} \vspace{-2em}
%\begin{align}\label{bath}
%W_b = \dfrac{1}{Z_b}&\int\mathcal{D}\phi_b\exp\left[\dfrac{1}{2}\int_0^\beta d\tau\int_{-a-\lambda}^{-a}dx\left[\left(\partial_\tau\phi\right)^2+\left(\partial_x\phi\right)^2\right]\right.\nonumber\\&\left.+\dfrac{1}{2}\int_{0}^{\beta}d\tau\int_a^{a+\lambda}dx\left[\left(\partial_\tau\phi\right)^2+\left(\partial_x\phi\right)^2\right]\right].
%\end{align}
\begin{align}\label{bath}
W_b=&\dfrac{1}{Z_b}\int\mathcal{D}\phi_b\exp\left[\dfrac{1}{2}\int_0^\beta d\tau\int_{-a-\lambda}^{-a}dx\left[\left(\partial_\tau\phi\right)^2+\left(\partial_x\phi\right)^2\right]\right.\nonumber\\&\qquad{}\left.+\dfrac{1}{2}\int_{0}^{\beta}d\tau\int_a^{a+\lambda}dx\left[\left(\partial_\tau\phi\right)^2+\left(\partial_x\phi\right)^2\right]\right].
\end{align}
%\vspace{-2em}\end{widetext}
Let $\xi=x+a$ for the first integral in Eq. (\ref{bath}) and
$\eta=x-a$ for the second. Fourier-expansion of the bath modes in
terms of the Matsubara frequencies \cite{Brink} gives us
%
%\begin{widetext} \vspace{-2em}
%\begin{subequations}\label{Fourier}
\begin{align*}
\phi(\eta,\tau)=&\dfrac{1}{\beta}\sum_m\left[\phi_m(a)\dfrac{\lambda-\eta}{\lambda}+\sum_{u=1}^\infty\phi_{um}\sin\left(\dfrac{\pi u\eta}{\lambda}\right)\right]e^{-i\nu_k\tau},\\
\phi(\xi,\tau)=&\dfrac{1}{\beta}\sum_m\left[\phi_m(-a)\dfrac{\lambda+\xi}{\lambda}+\sum_{u=1}^\infty\phi_{um}\sin\left(\dfrac{\pi u\eta}{-\lambda}\right)\right]e^{-i\nu_k\tau}.
\end{align*}
%\end{subequations}
%\vspace{-2em}\end{widetext}
%
$\nu_k = 2\pi k/\beta\;(k\in\mathbb{Z})$ are the bosonic Matsubara frequencies.
Substituting these expansions in Eq.
(\ref{bath}), we obtain
%\begin{widetext} \vspace{-2em}
\begin{align*}
W_b=\exp\left[\dfrac{1}{2\beta}\sum_m\left\lbrace|\phi_m(a)|^2 - |\phi_m(-a)|^2\right\rbrace\left(-\dfrac{\lambda\nu_k^2}{3}-\dfrac{1}{\lambda}+\sum_{u=1}^{\infty}\dfrac{2(\lambda\nu_k^2/\pi u)^2}{\lambda\nu_k^2+\pi^2u^2/\lambda}\right)\right]
\end{align*}
%\vspace{-2em}\end{widetext}
which in the limit $\lambda\to\infty$ gives 
\begin{align}
W_b=\exp\left[\dfrac{1}{2\beta}\sum_m|\nu_k|(|\phi_m(-a)|^2-|\phi_m(a)|^2)\right].
\end{align}
Therefore the effective generating functional (or reduced density matrix)
of the cavity fields is
%
%\begin{widetext} \vspace{-2em}
\begin{align}\label{reduceddensity}
W_c^R[\chi] =&\dfrac{1}{Z_c}\int\mathcal{D}\phi_c\exp\left[(\phi_m,\chi_{-m})+\dfrac{1}{2\beta}\sum_m\left[\int_{-a}^{a}dx\left[|\left(\partial_\tau\phi_m\right)|^2\right.\right.\right.\nonumber\\ &\qquad{}+|\left(\partial_x\phi_m\right)|^2 \left.\left.\left.+V(x)|\phi_m|^2\right]+|\nu_k|\left(|\phi_m(-a)|^2-|\phi_m(a)|^2\right)\vphantom{\int_{-a}^{a}}\right]\vphantom{\sum\limits_{min}^{max}\int_{-a}^{a}}\right].
\end{align}
%\vspace{-2em}\end{widetext}

\subsection{QNM expansion and path integral quantization}

We expand the cavity and the source fields in terms of the QNMs,
$\phi_m(x,0)=\sum_ma_{jm}f_j$ and
$\chi_{-m}(x,0)=\sum_mb_{j,-m}f_j$. We temporarily assume that
$V(x) = 0,\; x\in[-a,a]$. Therefore using the orthogonality condition
given by Eq.  (\ref{orthogon}) in Eq. (\ref{reduceddensity}) we obtain
%
%\begin{widetext} \vspace{-2em}
\begin{align}\label{cavitypath}
W_c^{R}[\chi]_0=&\exp\left[\dfrac{1}{8\beta}\sum_{jkm}\dfrac{b_{jm}b_{k,-m}}{\omega_j+\omega_k}\right.\left[f_j(-a)f_k(-a)\left(\dfrac{\theta(m)\omega_k}{i\omega_k+\nu_k}+\dfrac{\theta(-m)\omega_j}{i\omega_j-\nu_k}\right)\right.\nonumber\\&\left.\vphantom{\dfrac{1}{8\beta}\sum_{jkm}}\left.+f_j(a)f_k(a)\left(\dfrac{\theta(m)\omega_j}{i\omega_j+\nu_k}+\dfrac{\theta(m)\omega_k}{i\omega_k-\nu_k}\right)\right]\right].
\end{align}
%\vspace{-2em}\end{widetext}
%
We calculated this path integral by temporarily assuming that
$V(x)=0,\; x\in[-a,a]$.  Now for the actual case of $V(x)\neq0,\;
x\in[-a,a]$ we can calculate the cavity generating functional from
Eq. (\ref{cavitypath}) by perturbation techniques \cite{Brink}. Let
$a_{jm}\to\beta\partial/2\omega_j\partial b_{j,-m}$.  Then the
generating functional for the cavity is
%
%\begin{widetext} \vspace{-2em}
\begin{align}\label{finaldensity}
& W_c^R[\chi]\nonumber\\&=\dfrac{1}{Z_c}\exp\left[-\dfrac{\beta}{2}\sum_{m_1m_2}\sum_{j_1j_2}\int_{-a}^{a}dx\;V(x)f_{j_1}(x)f_{j_2}(x)\dfrac{1}{2\omega_{j_1}\omega_{j_2}}\dfrac{\partial}{\partial b_{j_1m_1}}\dfrac{\partial}{\partial b_{j_2m_2}}\vphantom{-\dfrac{\beta}{2}\sum_{m_1m_2}\sum_{j_1j_2}\int_{-a}^{a}dx}\right]W_c^R[\chi]_0
\end{align}
%\vspace{-2em}\end{widetext}
%
where we can substitute $V(x)=V_0$ for the rectangular barrier and $V(x) =
(1/4\nu)(\nu^2+c^2)\sech^2(\sqrt{\nu} x)$ for the modified
P\"{o}schl-Teller potential. Therefore Eq. (\ref{finaldensity}) gives
the generating functional of the cavity in terms of the QNMs of the
system only. As we have calculated the QNMs for our model potentials in Sec. (\ref{modelpot}),
we can just replace them in Eq. (\ref{finaldensity}) to obtain the
final algebraic forms of the generating functional.

\section{Dissipative oscillator models}\label{oscillators}
The Bateman-Feshbach-Tikochinsky (BFT) oscillator has been the inspiration behind our quantization process. However there's another oscillator system, the Caldirola-Kanai(CK) oscillator, which could have been used to model the quasi-normal modes. Here, we show that the BFT and the CK oscillators are equivalent. 
%Recently, Galley has come up with a new formulation for the nonconservative systems \cite{Galley2013}. One 
%of the crucial feature of Galley's approach is the formal doubling of the variables. While this formal 
%doubling of variables has been the feature for understanding dissipative oscillator, however, there is 
%no physical meaning associated to these extra variables. Here we show that using the field 
%theoretic concept it is possible to physically identify the doubling of modes in any dissipative system. 
%In specific, we use Galley's approach to obtain a one-to-one mapping between 
%the Bateman-Feshbach-Tikochinsky system and Caldirola-Kanai.

\subsection{Bateman-Feshbach-Tikochinsky System}Just like in the case of curved spacetime, a free scalar field in a flat spacetime is also governed by the Klein-Gordon equation
\begin{align}
	\label{Klein-Gordon-1}\left[\Box + m^{2}
	\right]\phi(\mathbf{x},t) = 0
\end{align} 
with $m$ being the mass of the field. Fourier transforming Eq. (\ref{Klein-Gordon-1}), in the scalar variables, we obtain
\begin{align}
	\label{KG-Fourier}	\ddot{y}_{\mathbf{k}}(t)+(k^{2}+m^{2})y_{{\mathbf{k}}}(t)=0
\end{align}
where 
$$ y_{{\mathbf{k}}}(t)\equiv\int\dfrac{d^{n}\mathbf{x}}{(2\pi)^{3/2}}e^{-i\mathbf{k}.\mathbf{x}}\phi(\mathbf{x},t). $$ 
Eq. (\ref{KG-Fourier}) is the equation of motion of a simple harmonic oscillator with the frequency being 
$\omega_{k}\equiv\sqrt{k^{2}+m^{2}}$. The scalar field can therefore be considered as a collection of an 
infinite number of simple harmonic oscillators with densely spaced frequencies. We now ask the question 
``What happens to the field if we include damping?'', i.e., what would be nature of the field decomposed 
in terms of oscillators instead of satisfying Eq. (\ref{KG-Fourier}):
$$
\ddot{y}_{\mathbf{k}}(t)+\omega_{k}^{2}y_{\mathbf{k}}+\alpha\dot{y}_{\mathbf{k}}=0
$$
where $\alpha$ is a spacetime resistance constant. To answer this question we start with a form invariant Lagrangian density. Using the Dirac matrices we construct a Lagrangian of the form
\begin{align}
	\label{New-Lagrangian}	\mathcal{L} = \dfrac{1}{2}\left(\partial_\mu\Phi^D\partial^\mu\Phi+\partial^\mu\Phi^D\partial_\mu\Phi\right)+\alpha\Phi^D\gamma^\mu\partial_\mu\Phi
\end{align} where $\Phi^D=\Phi^\dagger\gamma^0$ and $\Phi = (\phi_{1},\phi_{2},\ldots,\phi_{n})^{T}$ is a $n$-component spinor object. The $\gamma^\mu$'s are the Dirac matrices in $n$-dimensions.
%\begin{align}
%\label{New-Lagrangian}	\mathcal{L} = \dfrac{1}{2}\left(\partial_\mu\Phi^D\partial^\mu\Phi+\partial^\mu\Phi^D\partial_\mu\Phi\right)+\alpha(\mathbf{x},t)\Phi^D\gamma^\mu\partial_\mu\Phi
%\end{align} 
%
%where $\Phi^D=\Phi^\dagger\gamma^0$ and $\Phi = (\phi_{1},\phi_{2},\ldots,\phi_{n})^{T}$ is a $n$-component wavefunction, 
%with scalar fields as components. The $\gamma^\mu,\;\mu=0,1,2,\ldots,n$, are the Dirac matrices in $n$-dimensions. 
The Euler-Lagrange equation is
\begin{align}
	\label{fieldequationphi}
	\partial^{\mu}\partial_{\mu}\Phi-\alpha\gamma^{\mu}\partial_{\mu}\Phi=0.
\end{align}
For simplicity we work in $1+1$ dimensions, where the Euler-Lagrange equation reduces to a pair of coupled equations
%\begin{subequations}
\begin{equation}
	\begin{aligned}
		\partial_x^2\phi_1 - \partial_t^2\phi_1 - \alpha[\partial_t\phi_2 + \partial_x\phi_1] & =0\\
		%\end{align}
		%\begin{align}
		\partial_x^2\phi_2 - \partial_t^2\phi_2 - \alpha[\partial_t\phi_1 - \partial_x\phi_2] & =0.
	\end{aligned}
\end{equation}
%\end{subequations}
This is because the two dimensional Dirac matrices are
$$
\gamma^{0} = \begin{pmatrix}
0 & 1\\1 & 0
\end{pmatrix}\;\;
\text{and}\;\;
\gamma^1=\begin{pmatrix}
1 & 0\\0 & -1
\end{pmatrix}.
$$
If we naively assume a plane-wave like solution of the form  $ e^{i \omega x} (y_1(t),y_2(t))^T $, because of the first order position derivatives in the equation, we get a system of equations with complex parameters. But we can avoid this, thanks to the real $2 \times 2$ matrix representations of $\sqrt{-1}$, like for example $\sigma = \gamma^{0}\gamma^{1}$. Putting $\Phi = e^{\sigma \omega x} (y_1(t),y_2(t))^T$ in Eq. (\ref{fieldequationphi}) we obtain a coupled system of equations,
\begin{equation}
	\begin{aligned}
		\ddot{y}_{1}+\omega^{2}y_{1}+\alpha\dot{y}_2-\alpha y_1 & =0\\
		%\end{align}
		%\begin{align}
		\ddot{y}_{2}+\omega^{2}y_{2}+\alpha\dot{y}_1-\alpha y_2 & =0,
	\end{aligned}
\end{equation}
which we can easily decouple using $y_+=(y_1+y_2)/2$ and $y_-=(y_1-y_2)/2$. Thus, we obtain
\begin{equation}\label{bateman-dft}
	\begin{aligned}
		\ddot{y}_{+}+(\omega^{2}-\alpha\omega)y_{+}+\alpha\dot{y}_+ & =0\\
		%\end{align}
		%\begin{align}
		\ddot{y}_{-}+(\omega^{2}-\alpha\omega)y_{-}-\alpha\dot{y}_- & =0.
	\end{aligned}
\end{equation}
For $\omega > \alpha$, eqs.~(\ref{bateman-dft}) represent the Bateman dual system . Therefore we can consider the fields $\phi_{1}$ and $\phi_{2}$ to be
infinite collections of coupled oscillators - one of which is dissipative and the other is amplified.\\
\subsection{Caldirola-Kanai system from Galley's approach}
\label{Appendix}
Recently, Galley has come up with a new formulation for the nonconservative systems \cite{Galley2013}. One 
of the crucial feature of Galley's approach is the formal doubling of the variables. While this formal 
doubling of variables has been the feature for understanding dissipative oscillator, however, there is 
no physical meaning associated to these extra variables. Here we show that using the field 
theoretic concept it is possible to physically identify the doubling of modes in any dissipative system. 
Consider a simple harmonic oscillator with variable coordinate, $q(t)$ and natural frequency, $\omega$, coupled to N number of simple harmonic oscillators, $Q(\omega_n)$. The equations of motion are
\begin{align}\label{infinite-bath}
	M_n\ddot{Q}(\omega_n) + M_n\omega_n^2Q(\omega_n) &= \lambda(\omega_n)q\\\label{infinite-bath2}
	m\ddot{q}+m\omega^2q &= \sum_{n=1}^{N}\lambda(\omega_n)Q(\omega_n).
\end{align}
We double the degrees of freedom, $q\to(q_1,q_2)$ and $Q(\omega_n)\to(Q_1(\omega_n),Q_2(\omega_n))$ and define new coordinates 
$$q_\pm=\dfrac{q_1\pm q_2}{\sqrt{2}},\quad Q_\pm(\omega_n)=\dfrac{Q_1(\omega_n)\pm Q_2(\omega_n)}{\sqrt{2}}.
$$
Using the retarded and advanced Green's functions we write the equations of motion in terms of these new coordinates
\begin{align}
	Q_+(\omega_n,t) &= Q^h (\omega_n,t) + \dfrac{\lambda(\omega_n)}{M_n}\int_{t_i}^{t_f}dt'G_{r,n}(t-t')q_+(t')\\
	Q_-(\omega_n,t) &= \dfrac{\lambda(\omega_n)}{M_n}\int_{t_i}^{t_f}dt'G_{a,n}(t-t')q_-(t')
\end{align}
where $G_{r,n}(G_{a,n})$ is the retarded(advanced)Green's function of the $n^{\text{th}}$ oscillator and $Q^h(\omega_n)$ is a solution of the homogeneous equation. Using these solutions we can write the effective Lagrangian for the $q$-system as
\begin{align}\label{bateman-dual}
	L = m\dot{q}_+\dot{q}_- - m\omega^2q_+q_- + K(q_\pm,\dot{q}_\pm,t)
\end{align}
where $K(q_\pm,\dot{q}_\pm,t) = q_-\sum_{n=1}^{N}\lambda(\omega_n)Q^h(\omega_n,t)+\int_{t_i}^{t_f}q_-(t)G(t-t')q_+(t')dt'$ and $G(t-t')=\sum_{n=1}^{N}(\lambda_n^2)/(M_n\omega_n)\sin\omega_n(t-t')$. Let $Q^h(\omega_n,t)=0$, $M_n = M$ and $\lambda_n = \lambda(\omega_n)=\lambda\omega_n$. Then in the limit $N\to\infty$ and assuming that the frequencies are dense in $R_{+}$ we get
$$
G(t-t')=-\dfrac{\lambda^2}{M}\sum_n\dfrac{d}{dt'}\cos\omega_n(t-t')=-\alpha\dfrac{d}{dt'}\delta(t-t')
$$ where $\alpha$ is a constant. Therefore the effective Lagrangian becomes 
\begin{align}
	L = m\dot{q}_+\dot{q}_- - m\omega^2q_+q_- - \alpha\dot{q}_+q_-.
\end{align}
We define $x = (q_+-e^{\alpha t}q_-)/\sqrt{2}$ and $y = (q_-+e^{-\alpha t}q_+)/\sqrt{2}$. In terms of these new 
coordinates the effective Lagrangian becomes
%\vspace{-0.3em}
\begin{align}
	\label{eq:CK-Lag}
	L = m\dfrac{e^{\alpha t}}{2}(\dot{x}^2-\omega^2x^2) - m\dfrac{e^{-\alpha t}}{2}(\dot{y}^2-\omega^2y^2)+m \dfrac{d}{dt}\left(\dfrac{\lambda}{2}y^2e^{-\alpha t}\right).
\end{align} 
The Lagrangian in Eq. (\ref{eq:CK-Lag}) describes a dual-Caldirola-Kanai system. Therefore we see that the harmonic oscillator coupled to a bath of infinite 
number of oscillators (Eqs. (\ref{infinite-bath}) and (\ref{infinite-bath2})), the Bateman dual system (Eq. (\ref{bateman-dual}) and the Caldirola-Kanai system are all intimately connected.

\providecommand{\href}[2]{#2}\begingroup\raggedright\endgroup

%not cited references
%\nocite{Kokkotas}
%\nocite{Berti:2009kk}
%\nocite{Konoplya}
\nocite{Nollert5253}
\nocite{Skakala3}
\nocite{0264-9381-27-15-155004}
\nocite{Birmingham}
%\nocite{Beyer}
\nocite{Dekker198472}
\nocite{Ullersma196627}
\nocite{Yu}
\nocite{Grabert}
\nocite{Plank}
\nocite{Lai1987337}
\nocite{Brink2}
\nocite{Leung3152}

\end{document}